# Dynamic Shared Context Processing in an E-Collaborative Learning Environment

Jing Peng[1], Alain-Jérôme Fougères[1], Samuel Deniaud[1] and Michel Ferney[1]

[1]M3M Laboratory, University of Technology of Belfort-Montbéliard
Belfort, France
{jing.peng, alain-jerome.fougeres, samuel.deniaud, michel.ferney}@utbm.fr

**Abstract**
In this paper, we propose a dynamic shared context processing method based on DSC (*Dynamic Shared Context*) model, applied in an e-collaborative learning environment. Firstly, we present the model. This is a way to measure the relevance between events and roles in collaborative environments. With this method, we can share the most appropriate event information for each role instead of sharing all information to all roles in a collaborative work environment. Then, we apply and verify this method in our project with Google App supported e-learning collaborative environment. During this experiment, we compared DSC method measured relevance of events and roles to manual measured relevance. And we describe the favorable points from this comparison and our finding. Finally, we discuss our future research of a hybrid DSC method to make dynamical information shared more effective in a collaborative work environment.
***Keywords:*** *Dynamical Shared Context, Relevant Information Sharing, Computer Supported Collaborative Learning, CSCW*

## 1. Introduction

Everyone now recognizes that effective collaboration requires that each member of the collaboration receives relevant information on the activity of his partners. The support of this relevant information sharing consists of models for formalizing relevant information, processors for measuring relevance, and indicators for presenting relevant information. Various concepts and implementations of relevant information models have been presented in the areas of HCI and Ubiquitous Computing.

System Engineering (SE) is the application field of our research activity. According to ISO/IEC 15288:2002 standard [1], designing a system-of-interest needs 25 processes which are grouped into technical process, project processes, etc. The system life cycle stages are described by 11 technical processes. The first ones are the Stakeholder Requirements Definition Process, the Requirements Analysis Process and the Architectural Design Process. They correspond to the left side of the well-known entity Vee cycle [2] which links the technical processes of the development stage with the project cycle (described by the ISO/IEC 26702:2007 standard [3]). The system architect plays a significant role during the system development. Despite description of different tasks he has to realize, there is a risk of lack of understanding of his activity and of supporting the results of his system architecture design, in a collaborative manner.

Hence, to improve collaboration between system architects and other designers, we have defined two scientific objectives: 1) observing, studying and modeling the daily activity of system architect in order to produce new knowledge and to propose an activity model which describe his critical activity, and 2) facilitating cooperative activity of designers and system architects, in improving the sharing of their work context [4].

Regarding the second objective, we conducted two experiments in educational context (student projects for a course in digital factory):
- For the first experiment during spring semester 2009, we developed and implemented a computational model of relevant shared information (0-1 static model), based on an activity modelling using Activity Theory [5, 6] and a pre-acquisition of student interests, according to roles they would play in the proposed project. Kaenampornpan and O'Neill [5] have built a context classification model based on AT to define the elements of context. Cassens and Kofod-Petersen [6] have proposed the use of AT to model the context. By extension of this work, we proposed a context model consisting of subject context, object context, rule context, community context, tool context and division of labour context (Figure 1).
- For the second experiment, by repeating the same type of project during spring semester 2010, based on knowledge acquired from previous experience, we developed and implemented a dynamic computational model of relevant shared information (DSC model).

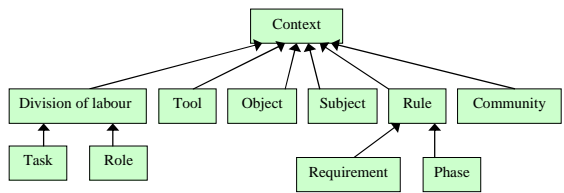

Fig. 1: Context modeling based on Activity Theory

In this paper, we firstly present DSC methodology which is a technique to measure the relevance between events and roles. Then, we apply this method in our collaborative environment project, and compare the related results with the results of manual-analyzing of the relevance. At last, we analyze the results of relevance measurement, and propose a new research direction to further improve the relevance of calculation of information sharing contexts.

## 2. Issue of this Research

Information technology has played an important role in the higher education infrastructure, and E-learning systems make the education more and more flexible and ubiquitous [7, 8]. Since learning is considered as a collaborative activity, a variety of CSCL (*Computer Supported Collaborative Learning*) systems have been developed in order to have a more effective study performance. With CSCL systems, students and teachers can perform all their activities in this virtual classroom environment, for instance, shared resources, disseminating knowledge and information, discussing, and following the project, etc.

In E-learning systems, sharing and disseminating information among the students and the teachers are very helpful and meaningful, such as learning resources, questions, feed-backs, etc. However in most CSCW (*Computer Supported Collaborative Work*) systems, sharing and disseminating information may cause a redundancy which has a bad influence on work and study performance [9, 10]. In our project, it is found that the redundant shared information will affect study motivation and performance. Therefore the elimination of the redundant shared and disseminated information becomes more and more crucial and important [11].

Measuring the relevance between events and roles, such as what the information retrieval does, when there is a query [12, 13]. In present study, that measurement is the way to do share more precisely, when an event occurs in the collaborative learning environment. We need to find the most appropriate roles or users to share this specific event information. This problem is resolved in three steps: firstly, how to capture and identify event attributes of an event when it is generated in the collaborative learning environment? Secondly, how to identify and evaluate a role's interest degree in the collaborative learning environment, which means: what kinds of events will be useful for him, and we can share these events information to him? Thirdly, how to measure the relevance degree between these roles and these events in the collaborative environment? In the following section, these steps will be explained in detail.

## 3. Method of Research

Our research has been carried out with two steps in two years experiments:
- *Step 1: Context Factor collection.* During the first experiment, we collected the context factors according to our context model from all the information of events.
- *Step 2: Relevance measurement.* During this experiment, we calculate the relevance dynamically through the context factor, which represent the event attribute.

DSC is a generic extensible context awareness method, which includes simple but powerful and lightweight mechanisms for the generation of relevant shared information to the appropriate users. The concept of DSC is based on event attributes, roles' interest degree, and relevance measurement. Relevance has been studied for the recent fifty years, and the general practice of relevance analysis is to research or to analyze the relationships between events or entities. The relevance measure or score is a quantitative result of the relevance analysis [14, 15, 16]. In this paper, our method is based on context model and relevance analysis (Figure 2). This method can be descried in three parts from left to right:
- *Part 1: Event capture and Role's interests capture.* In this part, we suppose that events could be presented by text and we gathered them during the experiment. Then we used Natural Language Processing tool to capture the key words as our context factors. Similar demonstration can be made for the role's interest capture. By supposing that the role's interests could be presented by text, we captured them during the ongoing of the experiment.
- *Part 2: Relevance measurement.* Details of this part are illustrated in the following paragraph.
- *Part 3: Relevantly shared information.* The measured information will be shared by relevance with different role in order to reduce the redundancy of information shared.

In this paper, we do not describe the first phase of this methodology, to focus on treatments carried out on knowledge (events and interests of roles). Simply put, the automatic capture of events, during the cooperative activity mediated by our environment, can be done automatically by an agent-based system [17, 18]. Indeed, agents can observe the work conducted in the environment and can capture and record all events and their context of appearance.

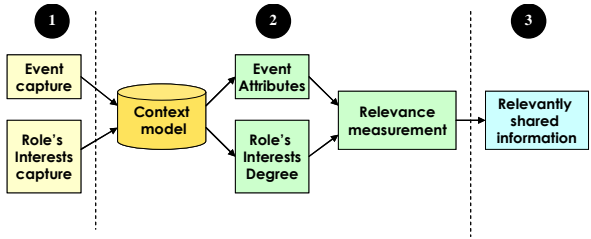

Fig. 2: Methodology of modeling Dynamical Shared Context (DSC)

### 3.1. Context Modeling

In the presented model, contexts are used to cluster information semantically. If we have a closer look at the term context, Wang [19] defines a context as "any information that can be used to characterize the situation of entities in the collaborative space." We propose a conceptual model that classifies contextual information into 7 categories as expressed in equation (1): Role context, Task context, Object context, Tool context, Requirement context, Community context and Event type context [5, 6, 20, 21, 22].
*Context = {Role, Object, Tool, Requirement, Community, Event_type}* (1)

In each category, there are many terms to present this context characteristic. We define these terms as *CF* (Context Factor): $c_i$ which can be a word, or a sentence. A *CF* is used to synthesize and describe an event attribute or a role's interest degree. In our case, *CF* is extracted by a probability method from a specific collaborative learning environment. We extract 60 CF from previous experiment, and we will use these CF in the next section.

Inspired from *TF-IDF* (Term Frequency – Inverse Document Frequency) method [23, 24, 25, 26], we define *IEF* (Inverse Event Frequency) as a measure of general importance of *CF* in a special corpus. Equation (2) presents the general importance of $c_i$ (*CF*):

$$ief_i = \log \frac{|E|}{|\{e : c_i \in e\}|} \quad (2)$$

with $|E|$ : total number of events in the corpus, $|\{e : c_i \in e\}|$ : number of events where $c_i$ appears.

With these *IEF* parameters of *CF*, event attributes and role's interest degree will be presented in following.

### 3.2. Event Attributes Modeling

For an event, we need a parameter to specify *EA* (Event Attributes), which is a set of importance degree of *CF* in a specific event. Firstly, we define *CFF* (Context Factor Frequency) in equation (3) to present the probability of $c_i$ in the total words appear in an event description or presentation.

$$cff_i = \frac{n_i}{\sum_k n_k} \quad (3)$$

where $n_i$ is the number of occurrences of the considered $c_i$ in an event (*W* is the sum of all the words in this event).

Then, we define *EW* (Event Weight) of $c_i$ in equation (4), which indicates the importance degree of $c_i$ in a specific event.

$$ew_i = cff_i \times ief_i \quad (4)$$

Finally, *EA* is obtained as (5):
$$ea = (ew_1, ew_2, ew_3, ... ew_n) \quad (5)$$

### 3.3. Role Interest Degree Modeling

A parameter presenting the *RID* (Role's Interest Degree), which represents the interest degree of *CF* for a specific role is needed. In the proposed method this can be presented and evaluated by *CF*. Firstly, we define role's *REF* (Relevant Event Frequency) in equation (6) to specify the probability of the events where $c_i$ appears in the total number of events relevant with a role.

$$ref_i = \frac{|\{e_r : c_i \in e_r\}|}{|E_r|} \quad (6)$$

with $|E_r|$ : total number of events in the corpus relative to this role, $|\{e_r : c_i \in e_r\}|$ : number of events relative to this role where $c_i$ appears.

Then, *RW* (Role Weight) of $c_i$ is defined in following equation for a role, which specifies the importance degree of $c_i$ for a specific role (7).

$$rw_i = ref_i \times ief_i \quad (7)$$

Finally, *RID* is obtained as (8):
$$rid = (rw_1, rw_2, rw_3, ... rw_n) \qquad (8)$$

### 3.4. Context Relevance Measurement

The relevance measurement is realized by cosine similarity method [27, 28, 29] in (9). Relevant sharing is a concept which describes the relevance of shared information. It considers that the relevance of shared information is a measurable and quantitative issue. In our opinion, the relevance of sharing is the most essential motive factor to share information in the collaborative work. Our aim is to measure relevance between events and roles.

$$relevance(v_e, v_r) = \cos\theta = \frac{v_e \cdot v_r}{\|v_e\| \|v_r\|} \qquad (9)$$

In practice, it is easier to calculate the cosine of the angle between the vectors. $\theta$ is the angle between event attributes vector and role's interest degree vector.

The event attributes vector for event *e* can be built from equation (10):
$$v_e = [ew_{1,e}, ew_{2,e}, ew_{3,e}, ... ew_{n,e}] \qquad (10)$$

The role's interest degree vector for role *r* can be built from equation (11):
$$v_r = [rw_{1,r}, rw_{2,r}, rw_{3,r}, ... rw_{n,r}] \qquad (11)$$

## 4. Experiment

In this section, we test the above method about calculating the relevance degree between events and roles by an example from our project. In practice, we apply this method in our Google App collaborative learning where it helps to send the email and announce to the appropriate receivers instead of all members in the project (Figure 3).

The project given to the students aims at designing an assembly system following the lean manufacturing principles, using the e-collaborative learning environment to manage cooperation in groups and project. The studied product is a hydraulic motor composed of 4 sub-assemblies (SA). The inherent structure of this product implies the use of the first three sub-assemblies to assemble the fourth sub-assembly (SA 4) and then creates the final product.

As shown in table 1, the students are divided into 2 groups of 15 persons. Each group aims at designing an assembly system based on a concept which is different from the one of the other group. The two different concepts of assembly system are firstly given to students at the beginning of the project. Then, each group is organized in 3 design teams: two teams are in charge of one sub-assembly (SA 1, SA 2) and the third one is in charge of the last sub-assembly (SA 4) and the whole assembly of the product. The other sub-assembly (SA 3) is assumed to be provided by an external provider.

Table 1: Project organization

| Product | Group 1 | Group 2 | Layer |
|---|---|---|---|
| *SA 1* | *Team 1* | *Team 1* | *L2* |
| *SA 2* | *Team 2* | *Team 2* | *L2* |
| *SA 3* | *External provider* | *External provider* | *L2* |
| *SA 4 +product* | *Team 3* | *Team 3* | *L2 + L1* |

The following roles are defined and assigned to team members respectively:
- Team manager (TM): Responsible of the planning and costs;
- Synthesis responsible (SR): Management of the architecture and simulation of the assembly system;
- Quality responsible (QR): Responsible of the quality according to the requirement and standard;
- Designer (DS), Common role to each team member.

According to the architectural Vee cycle [29], we identified two architectural layers. The top layer (L1) corresponds to the development of the whole assembly system and the bottom layer (L2) corresponds to the development of the three sub-assembly systems. Hence the team in charge of the whole assembly of the product is responsible of the top layer. From stakeholders' requirements, it defines the design requirements and architects the whole assembly system. Then it allocates to the bottom layer a set of derived requirements – the sub-assembly systems must satisfy. Finally, the synthesis responsible of the bottom layer provide the architectures of the sub-assembly systems which are then integrated into the whole assembly system by the synthesis responsible of the top layer in order to provide a system which verifies the synthesized stakeholders' requirements.

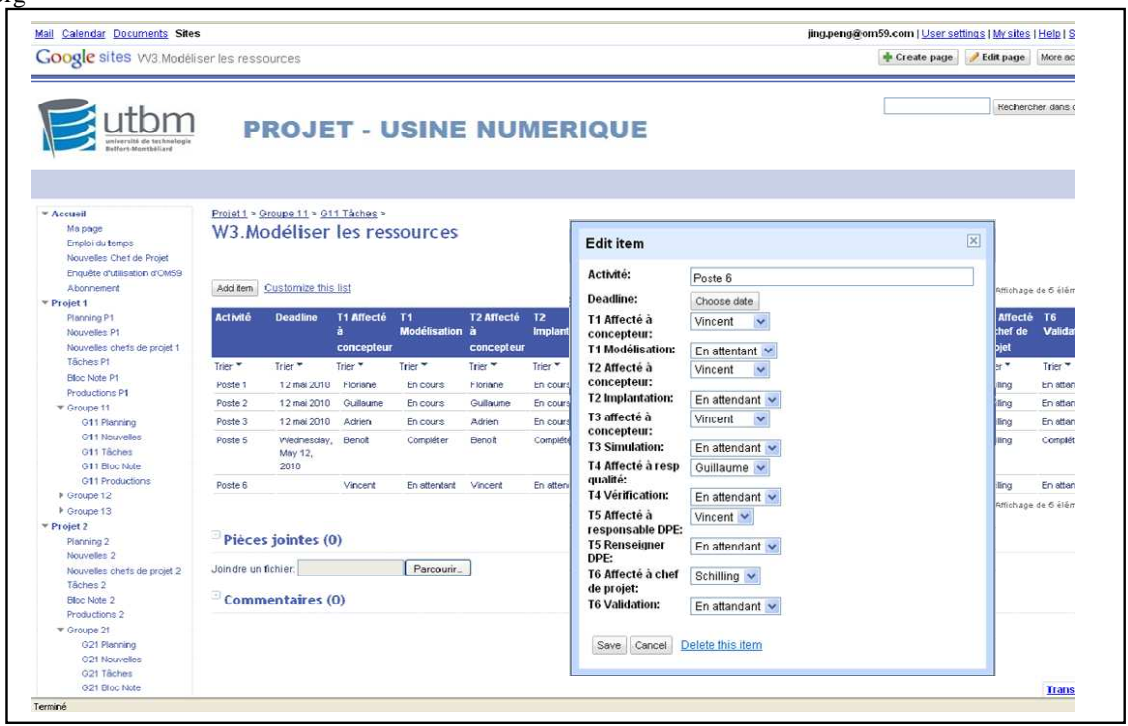

Fig. 3: Illustration of the e-collaborative learning environment used by the students: here, the project leader of Group 11 is currently editing the work of item 6 (information will be forwarded by email to members of Group 11)

### 4.1. Context Modeling

In our case, we get 219 samples of recorded events, such as email, message and question or notification. These context factors are classified according to our context model, and 82 typical event examples are selected to measure the relevance.

We take an example from *Team1* of *Group1* called *Group 11*. This event happened at the final step in the project cycle (Event 80, illustrate in Figure 4). The *Team1 Manager* announced their team's layout situation in digital factory relative to the other 2 teams. *Team1* Manager shared this information usually only with his team members, but this is not a relevant sharing by manual analysis. In this part, we measure relevance between this event and all the roles in *Group1* and compare calculating result of relevance with another model realized by a survey and manual-analysis of relevance. Finally, we discuss the meaning of results.

In this project, we calculated the *IEF* of all words in the events samples by using equation (2), and we selected the 64 highest *IEF* as our *CF* according to context model which are partially listed in table 2. The *CF* here was selected with context knowledge, not only with calculate results.

Table 2: Inverse Event Frequency

| CF | IEF |
|---|---|
| 1 | 1.4366926 |
| 2 | 0.83463261 |
| 3 | 0,46665582 |
| 4 | 0,95957134 |
| …… | ….. |
| 64 | 0,83463261 |

### 4.2. Event Attributes Modeling

In table 3, we present the number of each *CF* appeared in original description of this event example. Then we can calculate the *CFF* by dividing it by the total words in event by equation (3). There are 289 words in this event example.

Table 3: Context Factor Frequency

| CF | CFF |
|---|---|
| 1 | 0 |
| 2 | 0 |
| 3 | 0.00692042 |
| 4 | 0,00346021 |
| …. | ….. |
| 64 | 0 |

We present the *EW* of these context factors in table 4 calculated by equation (4).

Table 4: Event Weight

| CF | EW |
|---|---|
| *1* | *0* |
| *2* | *0* |
| *3* | *0,0031* |
| *4* | *0,0032* |
| *....* | *.....* |
| *64* | *0* |

The event attributes can be obtained via equation (12):
$ea=(0, 0, 0.0031, 0.0032, …… 0)$ (12)

Sequentially, we use the obtained event attributes to make the event attributes vector $v_E$ by equation (13)
$v_E = [0, 0, 0.0031, 0.0032, …… 0]$ (13)

### 4.3. Role Interest Degree Modeling

The different roles' *REF* calculated by equation (6) is listed in tables 5-7.

Table 5: Team1's roles Relevant Event Frequency

| CF | REF | | | | |
|---|---|---|---|---|---|
| | *TM* | *QR* | *SR* | *DS1* | *DS2* |
| *1* | 0.15 | 0 | 0.1428 | 0 | 0 |
| *2* | 0.2 | 0,3333 | 0.1428 | 0 | 1 |
| *3* | 0.75 | 0,3333 | 0.7143 | 0.6667 | 0 |
| *4* | 0.4 | 0,3333 | 0.1428 | 0 | 0 |
| *......* | *......* | *......* | *......* | *......* | *......* |
| *64* | 0.5 | 0 | 0.1428 | 0 | 0 |

Table 6: Team2's roles Relevant Event Frequency

| CF | REF | | | | |
|---|---|---|---|---|---|
| | *TM* | *QR* | *SR* | *DS1* | *DS2* |
| *1* | 0 | 0 | 0 | 0 | 0 |
| *2* | 0.1333 | 0,1111 | 0 | 0 | 0.5 |
| *3* | 0.2667 | 0,1111 | 0 | 0 | 0 |
| *4* | 0,0667 | 0 | 0 | 0 | 0 |
| *......* | *......* | *......* | *......* | *......* | *......* |
| *64* | 0,1333 | 0 | 0 | 0 | 0 |

Table 7: Team3's roles Relevant Event Frequency

| CF | REF | | | |
|---|---|---|---|---|
| | *TM* | *QR* | *SR* | *DS* |
| *1* | 0 | 0 | 0 | 0 |
| *2* | 0 | 0 | 0 | 0 |
| *3* | 0.1 | 0.125 | 0 | 0 |
| *4* | 0 | 0 | 0 | 0 |
| *......* | *......* | *......* | *......* | *......* |
| *64* | 0 | 0.125 | 0 | 0 |

The *RW* of these context factors calculated by equation (7) is listed in tables 8-10.

Table 8: Team1's Role Weight

| CF | RW | | | | |
|---|---|---|---|---|---|
| | *TM* | *QR* | *SR* | *DS1* | *DS2* |
| *1* | 0,2155 | 0 | 0,2052 | 0 | 0 |
| *2* | 0,1669 | 0,4789 | 0,1192 | 0 | 0,8346 |
| *3* | 0,3500 | 0,2782 | 0,3333 | 0,3111 | 0 |
| *4* | 0,3838 | 0,1556 | 0,1370 | 0 | 0 |
| *......* | *......* | *......* | *......* | *......* | *......* |
| *64* | 0,4173 | 0 | 0,1192 | 0 | 0 |

Table 9: Team2's Role Weight

| CF | RW | | | | |
|---|---|---|---|---|---|
| | *TM* | *QR* | *SR* | *DS1* | *DS2* |
| *1* | 0 | 0 | 0 | 0 | 0 |
| *2* | 0,1915 | 0,0927 | 0 | 0 | 0,4173 |
| *3* | 0,2226 | 0,0927 | 0 | 0 | 0 |
| *4* | 0,6077 | 0 | 0 | 0 | 0 |
| *......* | *......* | *......* | *......* | *......* | *......* |
| *64* | 0,1113 | 0 | 0 | 0 | 0 |

Table 10: Team3's Role Weight

| CF | RW | | | |
|---|---|---|---|---|
| | *TM* | *QR* | *SR* | *DS1* |
| *1* | 0 | 0 | 0 | 0 |
| *2* | 0 | 0 | 0 | 0 |
| *3* | 0,0467 | 0,0583 | 0 | 0 |
| *4* | 0 | 0 | 0 | 0 |
| *......* | *......* | *......* | *......* | *......* |
| *64* | 0 | 0,1043 | 0 | 0 |

The role *Team1* Manager's interest degree can be obtained by (14):
$rid_{T1M} = (0.2155, 0.1669, 0.3500, …, 0.4173)$ (14)

The *Team1* Manager's interest degree is sequentially used to make the role interest degree vector by equation (15):
$v_{T1M} = [0.2155, 0.1669, 0.3500, …, 0.4173]$ (15)

The same demonstration can be made to get other roles interest degree vectors, which are expressed in equations (16)-(28), respectively:
$v_{T1QR} = [0, 0.4789, 0.2782, 0.1556, …, 0]$ (16)
$v_{T1SR} = [0.2052, 0.1192, 0.3333, 0.1370, …, 0.1192]$ (17)
$v_{T1DS1} = [0, 0, 0.3111, 0, …, 0]$ (18)
$v_{T1DS2} = [0, 0.8346, 0, 0, …, 0]$ (19)
$v_{T2M} = [0, 0.1915, 0.2226, 0.6077, …, 0.1113]$ (20)
$v_{T2QR} = [0, 0.0927, 0.0927, 0, …, 0]$ (21)
$v_{T2SR} = [0, 0, 0, 0, …, 0]$ (22)
$v_{T2DS1} = [0, 0, 0, 0, …, 0]$ (23)

$v_{T2DS2}$ = [0, 0.4173, 0, 0, ..., 0]       (24)
$v_{T3M}$ = [0, 0, 0.0467, 0, ..., 0]       (25)
$v_{T3QR}$ = [0, 0, 0.0583, 0, ..., 0.1043]       (26)
$v_{T3SR}$ = [0, 0, 0, 0, ..., 0]       (27)
$v_{T3DS}$ = [0, 0, 0, 0, ..., 0]       (28)

### 4.4. Context Relevance Measurement

With the above mentioned method, we developed a simulator which can calculate the relevance degree between this event and all the roles in this project. This simulator uses DSC model and 0-1 model (mentioned in the following). It can also show the highest relevance members with this event sample in red, and this is a sharing recommendation for the user, when they want to share some information with others. The relevance measurement results are listed in tables 11-13.

A student survey conducted at the end of the project confirmed the interest and increasing relevance of information they had received over the project progressed. In this survey, we evaluated the students' satisfaction of relevance information shared recommended by our DSC model. To further improve the relevance of information to deliver, we now have to work on a hybrid model that combines the advantages of both models (TF-IDF model and 0-1 model): a previous knowledge acquisition (a priori interests expressed by the different roles), and a dynamic calculation based on events captured by the environment. Our first action in this direction has been to develop a tool to trace and compare the two approaches (Figure 4). This tool allows one to follow step by step the evolution of the dynamic calculation of relevance of information sharing for different roles, and to establish where the dynamic model TF-IDF is more efficient than the static model 0-1.

Table 11: Team1's roles Relevance

| Relevance | Team1 roles | | | | |
|---|---|---|---|---|---|
| | *T1TM* | *T1QR* | *T1SR* | *T1DS1* | *T1DS2* |
| DSC Model | 0.7654 | 0.3141 | 0.4218 | 0.2456 | 0.0405 |
| | High | High | High | Low | Low |
| 0-1 Model | 12 | 4 | 8 | 5 | 5 |
| | High | Low | High | Low | Low |
| Used to sharing | Yes | Yes | Yes | Yes | Yes |
| Manual analysis | Yes | Yes | Yes | No | No |

Table 12: Team2's roles Relevance

| Relevance | Team2 roles | | | | |
|---|---|---|---|---|---|
| | *T2M* | *T2QR* | *T2SR* | *T2DS1* | *T2DS2* |
| DSC Model | 0.4748 | 0.1224 | 0.1099 | 0.0551 | 0.0689 |
| | High | Low | low | Low | Low |
| 0-1 Model | 12 | 5 | 8 | 5 | 6 |
| | High | Low | High | Low | Low |
| Used to sharing | No | No | No | No | No |
| Manual analysis | Yes | No | No | No | No |

Table 13: Team1's roles Relevance

| Relevance | Team3 roles | | | |
|---|---|---|---|---|
| | *T3M* | *T3QR* | *T3SR* | *T3DS* |
| DSC Model | 0.4129 | 0.2465 | 0.2424 | 0.0912 |
| | High | Low | Low | Low |
| 0-1 Model | 12 | 4 | 8 | 5 |
| | Low | Low | High | Low |
| Used to sharing | No | No | No | No |
| Manual analysis | Yes | No | No | No |

From these results, we conclude that:

1. Relevance measured by DSC model shows that this event has high relevance with *T1M*, *T1QR*, *T1SR*, *T2M*, *T3M*, which are the same shared results by manual analysis.

2. In results from 0-1 model, some roles' relevance is not coherent with the manual analysis shared situation, such as *T1QR*.

3. In usually shared way, sharing information with which person is not accurate. By this sharing way, some roles have been missed, such as *T2M* and *T3M*; some roles have been shared with redundancy, such as *T1DS1* and *T1DS2*.

Through this example, we find that DSC model can make a reasonable relevance measurement between event and role relative to the manual analysis of sharing, and this application can make sharing more accurate and efficient.

Fig. 4: Tool for tracking and calculating relevance of information to share

## 5. Conclusion

E-collaborative learning systems have been developed to support collaborative learning activity, which makes the study more and more effective. Now we think that learning activity will have a better performance and effectiveness if we can eliminate the redundant shared information in the e-collaborative system. Therefore, we proposed an implementation of DSC method supported by an e-collaborative learning system environment. This environment can supply pertinent useful resources, propose appropriate workflow to realize project, share and disseminate relevant information.

We have described an original methodology of DSC model to address the problem of shared redundancy in collaborative learning environments. In this method, context model are applied for preparing the DSC modeling. A collective activity analysis approach (Activity Theory) allows us to build this context model. This phase is critical, since the dynamic shared context model depends on this context model to obtain satisfying context factors.

After describing our methodology we presented an illustration in an e-collaborative learning context. In this illustration, we selected 82 typical events from the project, and measured the relevance between these events and roles by using the tool Relevance Processing V_0.1. From the results of relevance, we got a useful advice to share the information in a collaborative environment.

To improve the relevance of information shared, we are now working in two directions: 1) proposition of a hybrid model, combining acquisition of interests as the roles of actors, and dynamic capture of events occurring in the course of a project, and 2) analysis of the relevance of a model based on a hierarchy of *CF*. This hierarchy could be established by an expert before start of collaborative projects.

## References

[1] ISO/IEC 15288:2002, "Systems engineering – System life cycle processes", ISO 2002.
[2] K. Forsberg, and H. Mooz, "The relationship of system engineering to the project cycle", in Proc. of the NCOSE conference, 1991, pp. 57-65.
[3] ISO/IEC 26702:2007, "Standard for Systems Engineering – Application and management of the systems engineering process", ISO 2007.
[4] J. Peng, A.-J. Fougères, S. Deniaud, and M. Ferney, "An E-Collaborative Learning Environment Based on Dynamic Workflow System", in Proc. of the 9th Int. Conf. on Information Technology Based Higher Education and Training (ITHET), Cappadocia, Turkey, 2010, pp.236-240.
[5] M. Kaenampornpan, and E. O'Neill, "Modelling context: an activity theory approach", In Markopoulos, P., Eggen, B., Aarts, E., Croeley, J.L., eds.: Ambient Intelligence: Second European Symposium on Ambient Intelligence, EUSAI 2004. pp. 367–374, 2004.
[6] J. Cassens, and A. Kofod-Petersen, "Using activity theory to model context awareness: a qualitative case study". In Proc. of the 19th International Florida Artificial Intelligence Research Society Conference, 2006, pp. 619-624.
[7] A. J. Cuthbert, "Designs for collaborative learning environments: can specialization encourage knowledge integration?", in Proc. of the 1999 conference on Computer support for collaborative learning (CSCL '99), International Society of the Learning Sciences, 1999.
[8] A. Dimitracopoulou, "Designing collaborative learning systems: current trends & future research agenda", in Proc. of the 2005 conference on Computer support for collaborative learning (CSCL '05), International Society of the Learning Sciences, 2005, pp. 115-124.
[9] L. Lipponen, "Exploring foundations for computer supported collaborative learning", in Proc. of the Conf. on Computer Support for Collaborative Learning (CSCL '02), International Society of the Learning Sciences, 2002, pp.72-81.
[10] A. Weinberger, F. Fischer, and K. Stegmann, "Computer supported collaborative learning in higher education: scripts for argumentative knowledge construction in distributed groups", in Proc. of the 2005 conference on Computer support for collaborative learning (CSCL '05), International Society of the Learning Sciences, 2005, pp. 717-726.
[11] T. Gross, and W. Prinz, "Modelling shared contexts in cooperative environments: concept, implementation, and evaluation", Computer Supported Cooperative Work 13,

Kluwer Academic Publishers, Netherlands, 2004, pp. 283-303.
[12] J. L. Dessalles, La pertinence et ses origines cognitives, Nouvelles théories, Hermes-Science Publications, Paris, 2008.
[13] R. Johnson, S. Harizopoulos, N. Hardavellas, K. Sabirli, I. Pandis, A. Ailamaki, N. G. Mancheril, and B. Falsafi, "To share or not to share?", in Proc. of the 33rd international conference on Very large data bases, Vienna, Austria, 2007, pp. 23-27.
[14] J. Mesmer-Magnus, and L. DeChurch, "Information sharing and team performance: A meta-analysis", Journal of Applied Psychology, Vol. 94, No. 2, 2009, pp. 535–546.
[15] M. Zhong, Y. Hu, L. Liu, and R. Lu, "A practical approach for relevance measure of inter-sentence", In Proc. of the 2008 Fifth international Conference on Fuzzy Systems and Knowledge Discovery, Vol. 4, 2008, pp. 140-144.
[16] W. B. Croft, and D. J. Harper, "Using probabilistic models of document retrieval without relevance information", Journal of documentation, Vol. 35, 1979, pp. 285-295.
[17] A.-J. Fougères, "Agents to cooperate in distributed design process", IEEE International Conference on Systems, Man and Cybernetics, (SMC'04), The Hague, Netherlands, 2004, pp.2629-2634.
[18] A.-J. Fougères, "Agent-Based µ-Tools Integrated into a Co-Design Platform", Int. J. of Computer Science Issues, Vol. 7, Issue 3-8, 2010, pp. 1-10.
[19] G. Wang, J. Jiang, and M. Shi, "Modeling contexts in collaborative environment: A new approach", Computer Supported Cooperative Work in Design III, 2007, pp 23-32.
[20] J. J. Jung, "Shared context for knowledge distribution: A case study of collaborative taggings", In Proc. of the 21st int. conference on Industrial, Engineering and Other Applications of Applied Intelligent Systems, New Frontiers in Applied Artificial Intelligence, Wrocław, Poland, 2008, pp. 641-648.
[21] A. Dey, D. Salber, and G. Abowd, "A conceptual framework and a toolkit for supporting the rapid prototyping of context-aware applications", Human-Computer Interaction Journal, Vol. 16 , No. 2-4, 2001, pp. 97-166.
[22] P. J. Brezillon, "Explaining for developing a shared context in collaborative design", in Proc. of the13th Int. Conf. on Computer Supported Cooperative Work in Design, 2009, pp.72-77.
[23] H. C. Wu, R. W. Luk, K. F. Wong, and , K. L. Kwok, "Interpreting TF-IDF term weights as making relevance decisions", ACM Trans. on Information System, Vol. 26, No. 3, 2008, pp. 1-37.
[24] T. Roelleke, and J. Wang, "TF-IDF uncovered: a study of theories and probabilities", In Proc. of the 31st Annual int. ACM Conf. on Research and Development in Information Retrieval (SIGIR '08), ACM, NY, 2008, pp. 435-442.
[25] K. Church, and W. Gale, "Inverse document frequency (idf): A measure of deviation from poisson", in Proc. of the third workshop on very large corpora, 1995, pp. 121-130.
[26] S. Tata, and J. M. Patel, "Estimating the selectivity of tf-idf based cosine similarity predicates". SIGMOD Rec. New York, Vol. 36, No. 2, 2007, pp. 7-12.
[27] M. Esteva, and H. Bi, "Inferring intra-organizational collaboration from cosine similarity distributions in text documents". In Proc. of the 9th ACM/IEEE-CS Joint Conf. on Digital Libraries, JCDL '09. ACM, New York, 2009, pp. 385-386.
[28] S. Yuan, and J. Sun, "Ontology-Based Structured Cosine Similarity in Speech Document Summarization", in Proc. of the 2004 IEEE/WIC/ACM int. Conf. on Web intelligence, IEEE Computer Society, Washington, 2004, pp. 508-513.
[29] K. Forsberg, H. Mooz, and H. Cotterman, Visualizing project management, models and frameworks for mastering complex systems, 3$^{rd}$ edition, John Wiley & Sons, 2005.

**Jing Peng** is Computer Engineer from University of Technology of Troyes (UTT). She is preparing PhD in Computer Science & System Engineering in the Laboratory of M3M at University of Technology of Belfort-Montbéliard (UTBM). Her research focus on context modeling, dynamical context shared and relevant information sharing.

**Alain-Jérôme Fougères** is Computer Engineer and PhD in Artificial Intelligence from University of Technology of Compiègne (UTC). He is currently member of the Laboratory of Mecatronics3M (M3M) at University of Technology of Belfort – Montbéliard (UTBM), where he conducts research on cooperation in design. His areas of interests and scientific contributions concern: the natural language processing, the knowledge representation, the multi-agent system design (architecture, interactions, communication and co-operation problems). In recent years, his research has focused on the context of co-operative work (mediation of cooperation and context sharing), mainly in the field of mechanical systems co-design.

**Samuel Deniaud**, Mechanical Engineer and holder of a PhD from the University of Franche-Comté, is presently Assistant Professor in the Mechatronic3M laboratory at the University of Technology of Belfort-Montbéliard (UTBM) since 2001. He conducts research activities in the field of system engineering, more particularly on the activities, methods and tools for designing system architecture from functional and dysfunctional points of view. He is currently applying his results on the architectural design activity of vehicles, and their associated production systems by using digital factory tools.

**Michel Ferney,** holder of a Ph.D. from the University of Franche-Comté (UFC), is presently Head Professor in the Laboratory of Mecatronics3M (M3M) at the University of Technology of Belfort – Montbéliard (UTBM). His research field of interest covers mechatronic product design and modeling, production systems modeling and soft computing.